\documentclass[iop,apjl]{emulateapj}
\usepackage{amsmath,amssymb,bm,amstext}

\usepackage{aas_macros}

\usepackage[breaklinks,colorlinks,citecolor=blue,linkcolor=magenta]{hyperref}

\begin{document}

\title{Fragmentation of Kozai--Lidov Disks}
\author{Wen Fu{$^{1,2,3}$}, Stephen H. Lubow{$^4$} and Rebecca G. Martin{$^1$}}
\affil{
{$^1$}Department of Physics and Astronomy, University of Nevada, Las Vegas, Las Vegas, NV 89154, USA\\
{$^2$}Department of Physics and Astronomy, Rice University, Houston, TX 77005, USA; wf5@rice.edu\\
{$^3$}Los Alamos National Laboratory, Los Alamos, NM 87545, USA\\
{$^4$}Space Telescope Science Institute, Baltimore, MD 21218, USA\\
}


\begin{abstract}
We analyze the gravitational instability (GI) of a locally isothermal
inclined disk around one component of a binary system. Such a disk can
undergo global Kozai--Lidov (KL) cycles if the initial disk tilt is
above the critical KL angle (of about $40^{\circ}$). During these
cycles, an initially circular disk exchanges its inclination for
eccentricity, and vice versa. Self--gravity may suppress the cycles
under some circumstances.  However, with hydrodynamic simulations
including self--gravity we show that for a sufficiently high initial
disk tilts and for certain disk masses, disks can undergo KL
oscillations and fragment due to GI, even when the Toomre $Q$ value
for an equivalent undisturbed disk is well within the stable regime
($Q > 2$).  We suggest that KL triggered disk fragmentation provides a
mechanism for the efficient formation of giant planets in binary
systems and may enhance fragmentation of disks in massive black hole
binaries.
\end{abstract}


\section{Introduction}
The stability of a self-gravitating gaseous disk is typically characterized by the Toomre  parameter, $Q=c_s \kappa/\pi G \Sigma$ \citep{Toomre64}, where $c_s$ is the sound speed, $\kappa$ is the epicyclic frequency, $G$ is the gravitational constant and $\Sigma$ is the disk surface density. The disk becomes unstable to gravitational instability (GI) when $Q$  is sufficiently small. This dynamical instability manifests itself as multi-armed spiral waves. The disk torques and shocks produced by these spiral structures redistribute mass and angular momentum, and ultimately help to stabilize the disk. With efficient radiative cooling or mass accretion onto the disk, GI can be maintained and leads to the collapse of regions on the spiral arms into gravitationally bound clumps. The exact condition for the development of disk GI into disk fragmentation is still an active area of research. Nevertheless, it is generally agreed that a necessary condition for GI against non-axisymmetric perturbations is $Q\lesssim 1.5$ \citep{Papaloizou91, Nelson98, Mayer04}. For non-isothermal disks, the $Q$ criterion is not sufficient for fragmentation because fragmentation depends also on the details of the disk thermodynamics \citep{Gammie01, Rice05, Lodato11, Paardekooper11,Kratter16}.

Disk GI has long been suggested to be an alternative theory for giant planet formation \citep{Boss97}. Compared with the more standard core-accretion theory, it has two advantages. First, massive planets form within a reasonable amount of time and secondly, it has the ability to form planets at large disk radii \citep[see review by][]{Helled14}. However this mechanism does not work close to the star where the disk can not cool efficiently \citep{Rafikov05} and the possibility of forming planets preferentially in the outer parts of the disk via disk GI has difficulty in explaining the majority of the known exoplanet population \citep{Rice15}.

There has been a plethora of numerical investigations (either with smoothed particle hydrodynamics codes or grid-based finite-difference/finite-volume code) devoted to studying various aspects of forming giant planets via disk GI, such as radiative transfer, effects of disk metallicity and chemistry, effects of numerical resolution and clump evolution \citep[e.g.][]{Boss09, Rogers12, Galvagni12, Nayakshin13, Vorobyov13, Stamatellos15, Evans15, Meru15, Tsukamoto15}. Most of these works assume that circumstellar disks form and exist in isolation. However, $40\%$ to $50\%$ of observed exoplanets are estimated to be in binary star systems \citep{Horch14}. 

Most models for GI involve a circular disk that surround a single star.
However, perturbations to the disk, such as those due to a binary companion, can potentially aid in producing fragmentation.
Some studies have investigated the possibility of forming giant planets by disk GI in binary systems, but they give conflicting conclusions regarding the role played by the binary companion in the disk GI development \citep{Nelson00, Mayer05, Boss06}. These papers modeled disks that are co--planar with the binary orbital plane. While disk co--planarity might be a good assumption under some circumstances, observational evidence suggests that it is not true for all binary systems. Large mutual inclinations (greater than $60^{\circ}$) have been observed between the circumstellar disks around young binary system components \citep[e.g.][]{Jensen14, Williams14}. The binary orbital planes in these systems are unknown but at least one of the disks in each system is possibly significantly inclined ( $> 45^\circ$) with respect to the binary orbit. Thus, it is important to understand disk GI and possible giant planet formation in inclined circumstellar disks. 

When the inclination of a tilted disk is greater than the critical KL angle (of about $40^{\circ}$), the disk can undergo global Kozai--Lidov (KL) oscillations \citep{Martin14, Fu15a}. The oscillations periodically exchange the disk eccentricity and inclination, as occurs in the ballistic particle case \citep{Kozai62, Lidov62}. In \cite{Fu15b}, we took a step further to include the effects of disk self-gravity. 
We found that the disk KL mechanism may be suppressed by disk self-gravity. However, the suppression ``window'' in terms of the disk mass and the disk inclination is quite narrow.  The disk KL mechanism
may still operate even when the disk in nearly gravitational unstable,
if the inclination is large.  

For disk tilts greater than about $60^o$, the KL effect attempts to induce a large
 disk eccentricity that cannot be fully accommodated by a smooth continuous disk. Instead,
 the disk undergoes strong shocks during its KL oscillations. Such shocks could have important consequences
 for disk fragmentation.

In this {\it Letter}, we report on finding of fragmentation triggered by the KL mechanism in the outer
disk regions.  In the current analysis, we assume that the disk is locally isothermal, as is more likely the case in outer disk regions, and ignore the complexities of nonisothermal effects. Our emphasis is on the dynamical
consequences of shocks driven in KL disks that provide much stronger disk perturbations than has been previously investigated
 \cite[e.g.,][]{Boss06}.
We describe our three-dimensional hydrodynamic simulation setup and main results in Section~2. We compare our study with earlier work on co--planar disks and discuss the implications of our results in Section~3. 

\section{Numberical Simulation}

\begin{deluxetable}{l c r}
\tabletypesize{\footnotesize}
\tablecolumns{3}
\tablecaption{SPH simulation parameters for an equal mass binary star system with total mass of $M$ and separation of $a_{\rm b}$\label{table:parameters}}
\tablehead{
\colhead{Binary and Disk parameters} & 
\colhead{Symbol} & 
\colhead{Value} 
}
\startdata
Mass of each binary component & $M_{\rm p}/M=M_{\rm c}/M$ & 0.5\\
Binary orbital eccentricity & $e_{\rm b}$ & 0\\
Initial number of particles & $N$ & $10^6$\\
Initial disk mass & $M_{\mathrm{disk}}/M$  & 0.035 \\
Initial disk outer radius & $r_{\mathrm{out}}/a_{\rm b}$ & 0.25\\
Initial disk inner radius & $r_{\mathrm{in}}/a_{\rm b}$ & 0.025\\
Mass accretion radius & $r_{\mathrm{acc}}/a_{\rm b}$ & 0.025\\
Disk viscosity parameter & $\alpha$ & 0.01\\
Disk aspect ratio & $H/r\, (r=r_{\mathrm{in}})$ & 0.1\\
Initial disk surface density, $\Sigma \propto r^{-\gamma}$ & $\gamma$ & 1.5\\
Initial disk inclination & $i_{\rm 0}$ & $\mathrm{70^{\circ}}$
\enddata
\label{tab:params}
\end{deluxetable}

We carry out three-dimensional smoothed particle hydrodynamical (SPH) simulations of a fluid disk  that orbits one member of an equal mass binary system. This binary system is initially on a circular orbit with separation $a_{\rm b}$. The total mass of the binary is $M=M_{\rm c}+M_{\rm p}$, where $M_{\rm c}$ is the mass of the central star and $M_{\rm p}$ is the mass of the perturber star.  The initial disk mass is $M_{\rm d} = 0.035 M$, or $7\%$ of the central stellar mass, $M_{\rm c}$. In our simulations, both the central star and the perturber star can feel the gravitational force from the disk and this affects the binary orbit. However, this effect is fairly small and the binary orbit remains almost circular throughout the whole simulation. Initially the orbital plane of the disk is inclined to the binary orbital plane by $i=70^{\circ}$ ($i=0$ would be a co--planar disk). We use a locally isothermal equation of state and an explicit accretion disk viscosity. The sound speed of the disk  is $c_{\rm s} \propto r^{-3/4}$ and the initial surface density distribution is $\Sigma \propto r^{-3/2}$. These are chosen such that both $\alpha$ \citep{SS73} and the smoothing length $\left<h\right> /H$ are constant over the disk radius, $r$ \citep{Lodato07}. We use $\alpha = 0.01$ in this study, and the disk initially extends from radius $r_{\rm in}=0.025\, a_{\rm b}$ to $r_{\rm out}=0.25 \,a_{\rm b}$.  The initial circular velocity is corrected for the effects of disk self-gravity. The vertical gas density distribution is taken to be that for hydrostatic balance of a nonself-gravitating disk. Although this initial state is somewhat out of vertical force balance due to self-gravity, this imbalance is unlikely to be responsible for the strong effects we find associated with large-scale shocks \citep[see also][]{Backus16}.  

We treat the stars as sink particles with a boundary condition such that whenever particles move into the accretion radius they are removed from the simulation while their mass and momentum are deposited onto the sink. 
There are $1 \times 10^6$ SPH particles at the beginning of the simulation. The disk aspect ratio at the inner disk edge is $H/R=0.1$ such that the shell-averaged smoothing length per scale height is $\left<h\right>/H\approx 0.26$ (i.e. the disk scale height is resolved by about $4$ smoothing lengths). The initial disk Toomre Q has a profile of $\sim r^{-3/4}$ such that the minimum $Q_{\rm min}\approx 2.2$, at $r=r_{\rm out}$ and $Q_{\rm max}\approx 12.4$, at $r=r_{\rm in}$. The simulation parameters are summarized in Table \ref{tab:params}. The parameters are very similar to those in \citep{Fu15b} except for the initial disk tilt. The dimensions of our parameters are all in 
length units of the binary separation, $a_{\rm b}$, mass units of the binary mass, $M=M_{\rm c}+M_{\rm p}$, and time units of the orbital period of the binary, $P_{\rm b}=2\pi \sqrt{a^3_{\rm b}/G(M_{\rm c}+M_{\rm p})}$. 
For instance, if $a_{\rm b}=100\,$AU and $M=1\,\rm M_{\odot}$, then both stars have mass of $0.5\,\rm M_{\odot}$ and the disk extends from $2.5\,$AU to $25\,$AU with a mass of $0.035\,\rm M_{\odot}$. Our simulation lasts for about $10$ binary orbits, which is about $10^4$ years. For a wider binary, the timescale for the simulation would be longer.

Our simulation tool is the {\sc phantom} code \citep{Lodato10, Price10, Price12, Nixon13}. We use a cubic spline kernel as the smoothing kernel. The number of neighbors is roughly constant at $N_{\rm neigh} \approx 58$. The viscosity follows the standard SPH prescription described in \cite{Monaghan92}. A viscosity switch is implemented to reduce the artificial viscosity away from shocks \citep{Balsara1995, Morris1997, Price04}. As is standard in SPH codes, we include a nonlinear term with a coefficient $\beta_{\rm AV}=2$ (AV stands for artificial viscosity) in order to suppress interparticle penetration.   The algorithm for the SPH implementation of self-gravity in {\sc phantom} is described in \cite{Price07b}, where the gravitational softening length is the same as the SPH kernel smoothing length \citep{Bate97}. The gravitation softening length is adaptive and is approximately equal to the SPH variable pressure smoothing length. To model disk fragmentation and clump formation, we use the sink particle creation feature of the code. This converts the gas particles near a local density maximum into a new sink particle.  The new sink particle does not include any gravitational softening. We follow \cite{Bate95} in choosing the conditions for new sink particle creation. These include checking whether the total mass within the kernel radius of the particle is at least one Jeans mass, whether the velocity divergence at the particle location is negative, and whether the material which will form the new sink has both thermal energy and rotational energy less than half of its gravitational energy. To avoid checking these criteria for every local density maximum which would greatly slowdown the computation, we set a critical density above which we implement these checks. This critical density is chosen to be $500\,Ma_{\rm b}^{-3}$ (note that the average disk density initially is about $38\,Ma_{\rm b}^{-3}$) and an accretion radius of $0.001\,a_{\rm b}$ is imposed around a newly-formed sink particle. If $a_{\rm b}=100\,$AU and $M=1\,\rm M_{\odot}$, then the critical density is $3\times10^{-10}$ $\mathrm{g/cm^3}$ and the accretion radius is $0.1\,$AU. 

\begin{figure*}
\centering
\includegraphics[width=0.7\textwidth]{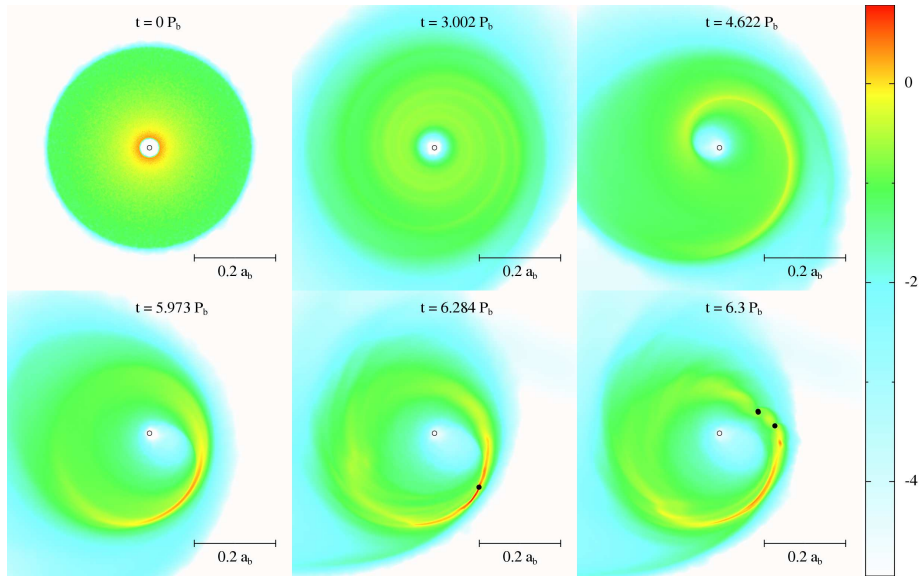}
\caption{Evolution of the self--gravitating disk with initial mass $M_{\rm d}=0.035\,M$, disk aspect ratio $H/R=0.1$ (at the inner disk radius), viscosity parameter $\alpha=0.01$ and initial inclination of $70^{\circ}$. Each panel corresponds to a  time (in units of the binary orbital period $P_{\rm b}$) and each panel has the same length scale. The disk is precessing and also changing tilt angle in time, but we show the face--on view of the disk. The binary separation is $a_{\rm b}$. The central binary component is at the center of each panel (denoted by a black circle) while the perturber star is out of the view on this scale. Newly formed clumps/fragments are denoted by black dots. The color coding is for the logarithm base $10$ of the column density (i.e. the density integrated along the line of sight) in units of $M/a^2_{\rm b}$. The disk initially is modeled with $10^6$ SPH particles. (Color online)}
\label{fig:fig1}
\end{figure*}

Figure \ref{fig:fig1} shows the evolution of the disk column density up to a time of about $6.3$ binary orbits.  
The disk  orientation changes due to both orbital precession and the KL oscillations but we always show the face-on view of the disk. Because of the spacial scale of the figure, the perturber star is out of the view, while the primary component is at the center of each panel. The top left panel shows that the disk is initially circular. After about $3$ binary orbits, the disk remains circular but the perturber star has driven spiral waves in the disk. In the top right panel which is at $t\approx 4.6\,P_{\rm b}$, the disk starts to show some eccentricity and a relatively strong one-arm spiral shock. At $t\approx 6\,P_{\rm b}$ (bottom left panel), the disk has clearly become more eccentric and the spiral shock has evolved into an arc-like shocked region with material being even more concentrated along that arc. Shortly after, at a time of $t\approx 6.28\,P_{\rm b}$ (bottom middle panel),  the first clump forms (denoted by the black dot). A second clump forms shortly after this as shown  in the last panel at time $t\approx 6.3\,P_{\rm b}$. Both clumps have mass of around $3\times10^{-5}\,M$. This makes them about $10\,\rm M_\oplus$  objects if we take the binary mass to be $M=1\,\rm M_\odot$. They form at a radius of $r\approx 0.2a_b$, where the local disk $Q$ value is now about $1.5$. By comparison, the local disk $Q$ value on the other side of the disk (also $\approx 0.2a_b$) is almost ten times higher ($Q\approx15$).

\begin{figure*}
\centering
\includegraphics[width=0.35\textwidth]{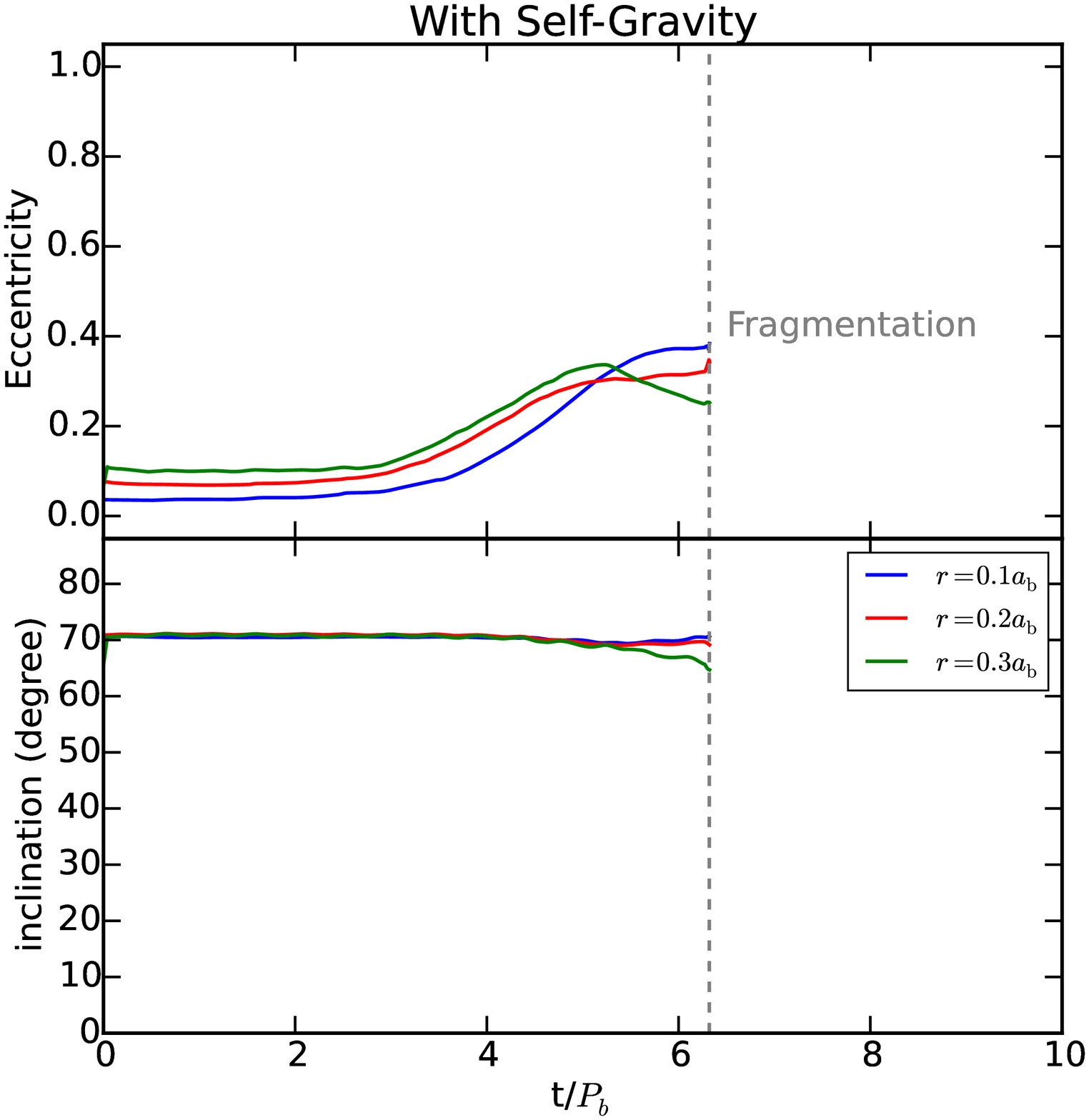}
\includegraphics[width=0.35\textwidth]{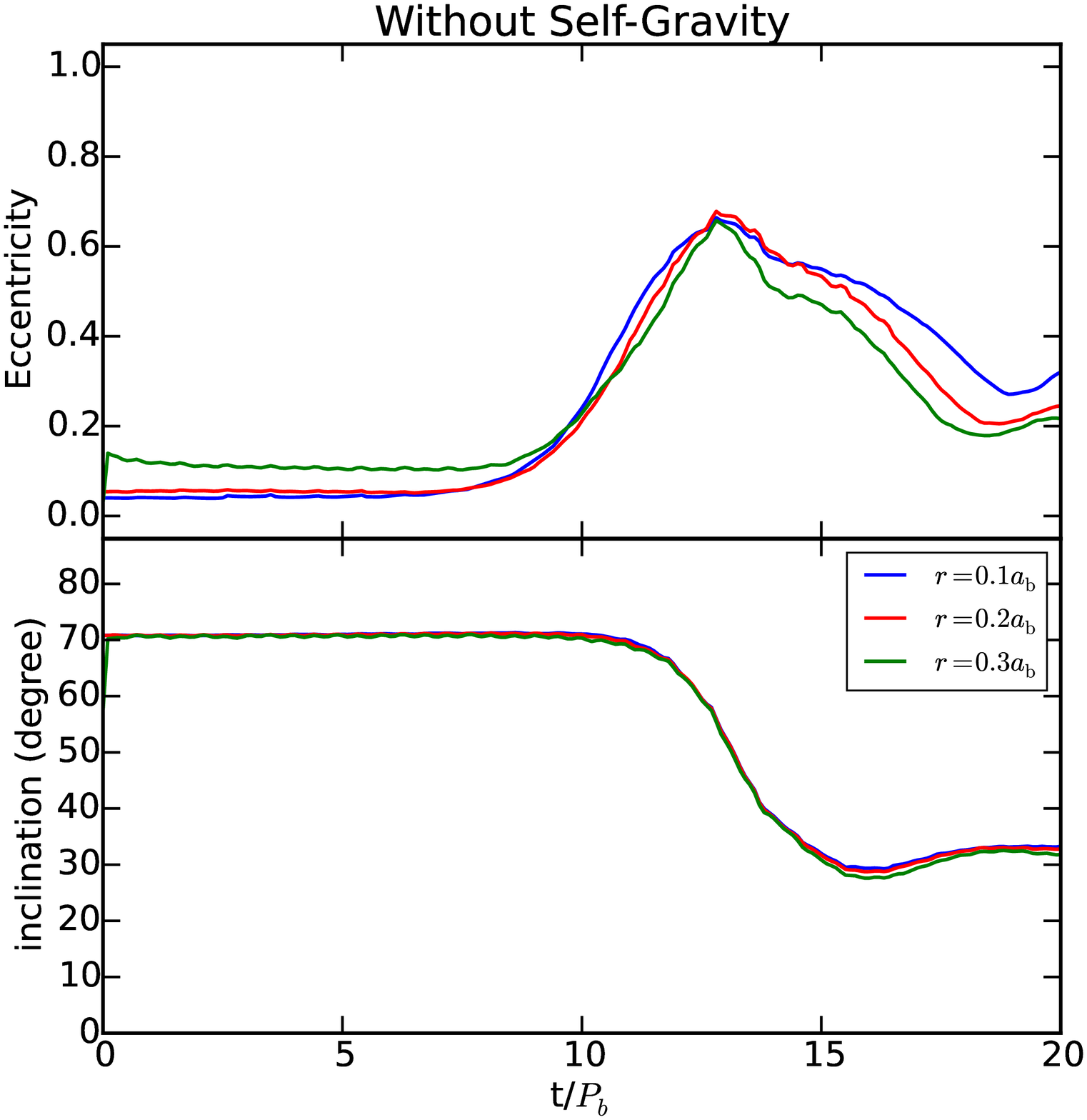}
\caption{Evolution of the eccentricity (upper panels) and inclination (lower panels) of the disk. We have averaged quantities over one binary orbit to smooth out small scale oscillations. The left panel shows the simulation including disk self-gravity (the run presented in Figure \ref{fig:fig1}) whereas the right panel shows the simulation without disk self--gravity. Different lines represent three different disk radii from the central star of $r=0.1\,a_{\rm b}$ (blue),   $r=0.2\,a_{\rm b}$ (red) and $r=0.3\,a_{\rm b}$ (green). Lines in the left column end at a time of about $t=6.4P_{\rm b}$ when disk fragmentation occurs. Lines in the right column exhibit the complete disk KL cycle when there is no self-gravity and therefore no fragmentation. (Color online)}
\label{fig:fig2}
\end{figure*}

In Figure \ref{fig:fig2} we show the time evolution of the eccentricity and inclination of the disk at three different radii from the central star. The left column shows the run presented in Figure \ref{fig:fig1} where disk self-gravity is included. The right column, in contrast, shows a simulation without disk self-gravity (and thus no disk fragmentation) while all the other simulation parameters are the same. The left column demonstrates that disk fragmentation occurs when the disk eccentricity has grown up to almost $0.4$. At this stage, the disk is still fairly early in its KL cycle so that the disk inclination has only declined slightly. If there is no fragmentation (see right column), then we see an interrupted disk KL oscillation during which disk eccentricity can grow up to about $0.7$ while the disk inclination falls to about $30^{\circ}$. Notice that in both simulations the disk remains quite flat throughout the simulation (see the similarity of disk inclination at the three different radii plotted). The KL cycle in the run including self--gravity starts at an earlier time than the one without self--gravity. This is because the initial driving of disk KL oscillation is very sensitive to initial model setup. Even slightly different initially conditions could lead to significantly different lengths of the pre-KL stage \citep[as discussed in][]{Fu15b}. Notice also that the disk eccentricity in the beginning is slightly higher than zero even though we set the disk up to be circular. This is due to the fact that we assume the SPH particles are on Keplerian orbits when computing their orbital eccentricities whereas they are actually on slightly sub-Keplerian orbits because of the disk pressure \citep{Fu15a, Fu15b}.

In \cite{Fu15b} we presented a simulation that is the same as the one in Figure~\ref{fig:fig1}, except that the disk initial tilt was $50^{\circ}$, as opposed to $70^{\circ}$ here. A disk with initial tilt of $50^{\circ}$ still undergoes KL oscillations, but the oscillation amplitudes are smaller (the peak disk eccentricity is about  $0.3$). There are two reasons for the weakened KL cycle in this case. First, the disk self-gravity acts to suppress the disk KL mechanism \citep{Fu15b}. Secondly, even without self-gravity, the disk KL oscillation amplitude is in general smaller for lower initial disk tilt \citep[see for example Figure 10 of][]{Fu15a}. In the $50^\circ$ inclination case, the disk still becomes eccentric  and looks similar to the first four panels of Figure \ref{fig:fig1}. However, the eccentricity growth is not strong enough for the shock front to be sufficiently dense to trigger disk fragmentation. We have also investigated two other cases that are not shown. We find that a  coplanar disk does not show disk fragmentation because the KL mechanism does not operate. Furthermore, for a lower disk mass, $M_{\rm d}=0.02\,M$, the disk $Q$ value is too high for fragmentation to occur. All of these results confirm that the disk fragmentation we see in Figure \ref{fig:fig1} is caused by disk GI that is triggered by strong eccentricity growth from disk KL cycle. It does not happen if the disk tilt is low which means either weak or completely no KL oscillation, at least for the disk mass ($7\%$ of the mass of the hosting star) we focus on in this study.

\section{Discussion and Conclusions}
In this {\it Letter}, we have shown that global KL oscillations in an inclined massive disk around one component of a binary system can facilitate the disk GI and fragmentation. For example, a disk of mass $0.035\,\rm M_{\odot}$ around a $0.5\,\rm M_{\odot}$ star can form multiple dense clumps of mass $10\,\rm M_{\oplus}$ on a timescale of about 6 binary orbits. This kind of disk fragmentation can occur even when the disk does not typically have a very small $Q$ value (e.g. $Q>2$). Thus, the same disk in isolation is fairly stable. It does require, however, that the disk initially has a high enough tilt so that a strong KL--driven disk eccentricity growth can be achieved. 

There are three previous papers that have simulated disk instability and fragmentation in binary star systems \citep{Nelson00, Mayer05, Boss06}. They reach different conclusions regarding whether or not a binary companion helps disk fragmentation in the co--planar disk case. The discrepancy of their results is probably due to differences in the simulation techniques (SPH code or grid-based code), numerical resolution, equation of state or binary orbit parameters \citep{Mayer10}.  Compared to these previous works, our disk is less massive (the minimum $Q\approx 2.2$) and quite stable against GI without the binary companion. The disks in these three papers are just marginally stable   (the minimum $Q\sim 1.3-1.9$). In these papers, it generally takes a few hundred years to see disk fragmentation, if there is indeed binary companion induced disk instability. In our simulation, the disk fragments after it has gained enough eccentricity. In these papers, the disk is still quite circular at the time of fragmentation. 

We have taken the disk to behave isothermally. This condition
makes fragmentation occur more easily than it would in a disk that responds adiabatically due to increased resistance to compression
\citep{Gammie01, Rafikov05}.   Self-gravitating disks in binaries can
be stabilized against fragmentation as a result of the extra
heating associated with the perturbation \citep[e.g.,][]{Mayer10}.
Therefore,  possible nonisothermal effects are important to  include in the analysis.

 We have adopted a binary separation of 100AU. A wider binary of a few hundred AU or more would permit the disk to be truncated
at larger radii \citep{AL94}, greater than about 40 AU, where a disk is more amenable fragmentation because the optical depth is closer to unity \citep[e.g.,][]{Hayfield11}. 
The KL oscillation timescale would be longer in a wider binary, but  could still shorter than the disk lifetime.

Because our code simply treats newly formed clumps as sink particles, we are not able to study the internal evolution of these dense objects and interactions between sink particles (e.g. merging of two clumps). In our study, we intentionally stop the simulation shortly after the formation of the first one or two clumps, even though the code can still proceed with formation of even more clumps. It is not our goal to follow the full path of the formation, evolution of any disk clump. That requires a much more sophisticated numerical investigation which is beyond the scope of this study.
Whether these disk fragments can remain as gravitationally bound objects and  what their final distributions are deserve to be the subject of future investigations. 

We have concentrated on KL fragmentation of protostellar disks. Other disk environments that are subject to fragmentation
around a single object may experience enhanced fragmentation conditions in a noncoplanar binary environment through the KL effect. For example,
fragmentation may occur in a disk around  a massive black hole \citep{Goodman03, Rafikov09}.
KL-induced shocks may expedite such fragmentation involving  binary black hole systems with small separations where the binary mass dominates.


W.F. and S.H.L. acknowledge support from NASA grant NNX11AK61G.  Computing resources supporting this work were provided by the institutional computing program at Los Alamos National Laboratory. We thank Daniel Price for providing the {\sc phantom} code for SPH simulations and the {\sc splash} code \citep{Price07a} for data analysis and rendering of figures.


\end{document}